\documentclass[onecolumn,nobibnotes,nofootinbib]{revtex4}
\usepackage{amsmath,amssymb}
\usepackage{graphicx,subfigure,epsfig}

\newcommand{\be}{\begin{equation}}
\newcommand{\ee}{\end{equation}}
\newcommand{\bea}{\begin{eqnarray}}
\newcommand{\eea}{\end{eqnarray}}
\newcommand{\benn}{\begin{displaymath}}
\newcommand{\eenn}{\end{displaymath}}
\newcommand{\beann}{\begin{eqnarray*}}
\newcommand{\eeann}{\end{eqnarray*}}

\begin{document}

\title{On possible implications of gluon number fluctuations in DIS data}
\author{Misha Kozlov}
\email{mkozlov@physik.uni-bielefeld.de}
\author{Arif Shoshi}
\email{shoshi@physik.uni-bielefeld.de}
\author{Wenchang Xiang}
\email{wcxiang@physik.uni-bielefeld.de} \affiliation{Fakult\"at f\"ur Physik, 
Universit\"at Bielefeld, D-33501 Bielefeld, Germany}

\begin{abstract}
  We study the effect of gluon number fluctuations (Pomeron loops) on deep
  inelastic scattering (DIS) in the fixed coupling case. We find that the
  description of the DIS data is improved once gluon number fluctuations are
  included. Also the values of the parameters, like the saturation exponent
  and the diffussion coefficient, turn out reasonable and agree with values
  obtained from numerical simulations of toy models which take into account
  fluctuations. This outcome seems to indicate the evidence of geometric
  scaling violations, and a possible implication of gluon number fluctuations,
  in the DIS data. However, we cannot exclude the possibility that 
  the scaling violations may also come from the diffusion part of the
  solution to the BK-equation, instead of gluon number fluctuations.



%
\end{abstract}

\maketitle    

\section{Introduction}
\label{sec:intro}

The mean-field dynamics of the high-energy dipole-proton scattering is
described by the BK-equation~\cite{Kovchegov:1999yj+X}. Phenomenological
ans\"atze for the dipole-proton scattering amplitude $T(r,x)$ (where $r$ is
the transverse dipole size and $x$ the Bjorken-variable) inspired by the
BK-equation have led to quite successful descriptions of the HERA data. The
$T$-matrix following from the BK-equation shows within a restricted
kinematical window, which increases with collision energy, the geometric
scaling behaviour~\cite{Mueller:2002zm,Munier:2003vc,Iancu:2002tr}, $T(r,x)
=T(r^2\,Q_s^2(x))$, where $Q_s(x)$ is the saturation scale, which seems well
supported by the HERA data~\cite{Stasto:2000er}.  The correction to the
solution outside the restricted window, the ``BK-diffusion term'', violates
the geometric scaling~\cite{Mueller:2002zm,Munier:2003vc,Iancu:2002tr} and
depends on the variable $\ln(1/r^2\,Q_s^2(x))/\sqrt{D_{\mbox{\footnotesize BK}}Y}$.
Iancu, Itakura and Munier (IIM)~\cite{Iancu:2003ge} have shown that the
``BK-diffusion term'', giving a substantial amount of geometric scaling
violations, is needed in order to accurately describe the experimental HERA
data. The exponent $\lambda$ of the saturation scale, $Q^2_s(x) \simeq
(x_0/x)^\lambda$, is known at NLO~\cite{Triantafyllopoulos:2002nz},
$\lambda\simeq 0.3$, and agrees with the values extracted from fits to HERA
data.



Recently, there has been a tremendous theoretical progress in understanding
the high-energy QCD evolution beyond the mean field approximation, i.e. beyond
the BK-equation. It has been understood how to include discreteness and
fluctuations of gluon numbers (Pomeron loops) in small-$x$
evolution~\cite{Mueller:2004se,
  Iancu:2004es,Mueller:2005ut,Iancu:2004iy,Kovner:2005nq}. After including
these elements, the evolution becomes stochastic and one has to distinguish
between the event-by-event amplitude $T(r,x)$, which corresponds to an
individual gluon number realization, and the physical amplitude $\langle
T(r,Y)\rangle$, which one obtains by averaging over all individual
realizations~\cite{Iancu:2004es}. At very high energy, the discreteness effect
{\em decreases} the exponent $\lambda$ as compared to BK-value and the gluon
number fluctuations, i.e., the averaging over all events to calculate the
physical amplitude, replaces the geometric scaling resulting from the
BK-equation (in the ``wave front'' region) by a new
scaling~\cite{Mueller:2004se, Iancu:2004es}, the {\em diffusive scaling},
namely, $\langle T(r,Y)\rangle$ is a function of a single variable
$\ln(1/r^2\,Q_s^2(x))/\sqrt{D\,Y}$, where $D$ is the diffusion coefficient.
The value of $D$ determines the rapidity above which gluon number fluctuations
become important, $Y \geq Y_{D}=1/D$, which is the case when the fluctuation
of the saturation scales of the individual events becomes large, in formulas,
when the dispersion $\sigma^2 = 2 (\langle \rho^2_s(Y)\rangle - \langle
\rho_s(Y)\rangle^2) = D\,Y \gg 1$, where $\rho_s(Y) = \ln(Q^2_s(Y)/Q_0^2)$. At
high energy, such that $\sigma^2 \gg1$, it has been shown that fluctuations do
strongly modify measurable quantities~\cite{Kozlov:2006qw,Iancu:2006uc}.  (A
more detailed presentation of the recent theoretical progress is given in
Refs.\cite{Mueller:2005me+X} while the most recent studies on Pomeron loops
based on toy models can be found in
Refs.~\cite{Shoshi:2005pf,Bondarenko:2006rh,Blaizot:2006wp,Iancu:2006jw,Dumitru:2007ew,Munier:2006um,Kozlov:2006zj,Kozlov:2006cu,Levin:2007yv}.)

In this work we elaborate, in a quite approximative way, whether the HERA
data~\cite{Breitweg:2000yn+X} do indicate any possible implication of gluon
number fluctuations. The coupling is kept fixed throughout this work. We
proceed in the following way: We use for the event-by-event amplitude several
models, the GBW model~\cite{GolecBiernat:1998js}, the IIM
model~\cite{Iancu:2003ge} and a model which is close to the theoretical
findings for $T$ at very large energy (see Eq.~(\ref{eq:ebe})).  For the
averaging over all events we use the high-energy QCD/statistical physics
correspondence~\cite{Iancu:2004es}, i.e., a Gaussian for the distribution of
$\rho_s(Y) = \ln(Q^2_s(Y)/Q_0^2)$. Moreover, assuming that the DIS cross
section shows diffusive scaling in the HERA energy range, we have used the
``quality factor'' method of Ref.~\cite{Gelis:2006bs} to get an estimation for
the value of $\lambda$, in a model-independent way. The procedure we use in
this work is always based on approximations and, therefore, can at best give
hints on a possible implication of gluon number fluctuations in the HERA data.

After including fluctuations in the way described above, we obtain from the
analysis of the HERA data values for the exponent $\lambda$ and the diffusion
coefficient $D$ which are quite independent of the ansatz for the
event-by-event amplitude. Also the model-independent approach gives a similar
value for $\lambda$. We find $\lambda \simeq 0.2$ which is smaller
than the value from the BK-inspired models (no fluctuations), $\lambda\simeq 0.3$,
and the decrease is in agreement with theoretical expectations. For the
diffusion coefficient we find a sizeable value, $D \simeq 0.35$. Surprisingly,
this value is very close to the values found for $D$ in numerical simulation
of the $(1+1)$ dimensional model~\cite{Iancu:2006jw} and of evolution
equations in QCD~\cite{Soyez:2005ha} (approximations to Pomeron loop
equations~\cite{Mueller:2005ut,Iancu:2004iy,Kovner:2005nq}) in the fixed
coupling case. The sizeable value of $D$ may indicate a possible involvement
of fluctuations in the HERA data since $Y \geq Y_D=1/D$ for rapidities at
HERA.

We observe that after including fluctuations the description of the HERA data
is improved for all models we have used for the event-by-event amplitude.  In
the case of the GBW model, which exhibits pure geometric scaling, after the
inclusion of fluctuations, which lead to a violation of geometric scaling, a
much better description is obtained, namely, $\chi^2/\mbox{d.o.f} = 1.74$
without and $\chi^2/\mbox{d.o.f} = 1.14$ with fluctuations. The situation
seems to be similar with all event-by-event amplitudes which show geometric
scaling.  In the case of the IIM model, which contains already the geometric
scaling violating BK-diffusion term, the inclusion of fluctuations also
improves, however less than in the GBW case, the description of the HERA data;
$\chi^2/\mbox{d.o.f} = 0.983$ before and $\chi^2/\mbox{d.o.f} =
0.807$ after
including fluctuations.\footnote{The $\chi^2$ is defined such that the smallest
  $\chi^2$ gives the best description to the HERA data.}  The outcomes seem to tell us that violations of
geometric scaling are required for an accurate description of the HERA data.
The improvement of the description of the HERA data together with the very
reasonable values for the parameters discussed above seem to indicate that
gluon number fluctuations may be the reason for geometric scaling violations
in the HERA data. However, we wish to emphasize here that the BK-diffusion
term gives similar geometric scaling violations as fluctuations and may as
well be the reason for the geometric scaling violations in the HERA data.

This work is organized as follows: In Sec.~\ref{sec:mfa}, we show the results
for the $T$-matrix for dipole-proton scattering and for the energy dependence
of the saturation scale which are obtained in the mean field approximations,
i.e., from the BK-equation. The results for the same quantities beyond the
mean field approximation, or the effects of discreteness and fluctuations in
gluon numbers on these quantities, are summarized in Sec.~\ref{sec:bmfa}.
Finally, we give numerical results and discuss a possible implication of
the physics beyond the mean field approximation in the HERA data.

\section{Mean field approximation} 
\label{sec:mfa}
In the mean field approximation, the $Y$-dependence of the $T$-matrix for a
dipole of transverse size $r$ scattering off a proton is given by the
BK-equation. In the fixed coupling case, the
solution to the BK-equation in the saturation region, where $T \simeq 1$,
is~\cite{Levin:1999mw}
%
%
\be 
T(r,Y) = 1- C_0\ \exp\left[-C_1 (\rho-\rho_s(Y))^2\right]\ \quad
\mbox{for} \quad \rho-\rho_s(Y) \ll 1 \
, 
\ee
while for the front of the $T$-matrix, where $T \ll 1$ (but not too small), one finds~\cite{Mueller:2002zm,Munier:2003vc}
\be
T(r,Y) = C_2 \ \left[\rho-\rho_s(Y)+C_3\right]\ 
\exp\left[-\lambda_s (\rho-\rho_s(Y)) -\frac{(\rho-\rho_s(Y))^2}
                {2 \bar{\alpha} \chi''(\lambda_s)Y}\right] \quad \mbox{for} \quad
            1 \ll \rho-\rho_s(Y) \ll  2 \chi^{{\prime\prime}}(\lambda_s) 
\bar{\alpha}_s Y \  \ , 
\label{eq_T_fQ}
\ee
where have used $\bar{\alpha}_s = \alpha_s N_c/\pi$, $\rho =
\ln(1/r^2\,Q_0^2)$ and $\rho_s(Y)=\ln(Q^2_s(Y)/Q_0^2)$ with $Q_s(Y)$ the
saturation scale. In above equations, the constants $C_0$, $C_2$, $C_3$ are of
${\cal O}(1)$, $C_1 = -C_F(1-\lambda_0)/N_c 2 \chi(\lambda_s)$ ($C_F$ is the
casimir factor in the funcdamental respresentation), $\lambda_s =
0.6275$, and $\chi(\lambda) = 2 \psi(1)-\psi(\lambda)-\psi(1-\lambda)$ is the
eigenvalue of the BFKl kernel. For the rapidity dependence of the saturation
scale, which separates the saturated ($r \gg 1/Q_s(Y)$) from the dilute ($r
\ll 1/Q_s(Y)$) regime, one obtains from the BK
equation~\cite{Mueller:2002zm,Munier:2003vc}
\be
Q^2_s(Y) = Q^2_0\ \frac{\exp\!\left[\bar{\alpha}\chi'(\lambda_s)Y\right]}
     {\left[\bar{\alpha} Y\right]^{\frac{3}{2
           (1-\lambda_0)}}} \ .
\label{eq_Q_s}
\ee
Note that within the even more restricted window, $\rho-\rho_s(Y) \ll
\sqrt{2 \chi^{^{\prime\prime}}(\lambda_s) \bar{\alpha}_s Y}$, where the
diffusion term in the exponent in Eq.(\ref{eq_T_fQ}) can be neglected, the
$T$-matrix shows the geometric scaling behaviour, i.e., it depends only
on the difference $\rho-\rho_s(Y)$ intead of depending on $r$ and $Y$
separately. At very large $r$, so that $\rho-\rho_s(Y) \gg  2
\chi^{{\prime\prime}}(\lambda_s)\bar{\alpha}_s Y$, the $T$-matrix exhibits color
transparency, i.e., it shows a faster decrease with $\rho$ as
compared to Eq.(\ref{eq_T_fQ}); $T \sim \exp[-\rho]$.  

Iancu, Itakura and Munier~\cite{Iancu:2003ge} have used the following ansatz for
the $T$-matrix,
\begin{equation}
T^{\mbox{\footnotesize IIM}}(r,Y) = \left\{ \begin{array}{r@{\quad,\quad}l}
1- \exp\left[ - a \ln^2(b\,r\,Q_s(x))\right] & r\,Q_s(x) > 2
\vspace*{0.5cm}\\
N_0 \left(\frac{r\,Q_s(x)}{2}\right)^{2\left(\lambda_s +
    \frac{\ln(2/r\,Q_s(x))}{\kappa\,\lambda\,Y}\right)} & r\,Q_s(x) <
  2  \ ,
\end{array} \right.
\label{eq:IIM}
\end{equation}
which obviously includes the features of the solution to the BK equation, to
compare the theory in the mean field approximation with the DIS data.  They
have used for the saturation momentum the leading $Y$-dependence of
Eq(\ref{eq_Q_s}), $Q_s(x) = (x_0/x)^\lambda$, however, with $\lambda$ and
$x_0$ being fixed by fitting the DIS data. The constant $\kappa =
\chi^{\prime\prime}(\lambda_s)/\chi^{\prime}(\lambda_s) \approx 9.9$ is a LO
result coming from the BK-equation, $N_0$ is a constant around $0.5$ and $a$
and $b$ are determined by matching the two pieces in Eq.(\ref{eq:IIM}) at
$r\,Q_s =2$.

The ``BK-diffusion term'' in the IIM-ansatz~(\ref{eq:IIM}),
\be 
\left(\frac{r\,Q_s(x)}{2}\right)^{2
  \frac{\ln(2/r\,Q_s(x))}{\kappa\,\lambda\,Y}} =
\exp\left[-\frac{\ln^2(4/r^2\,Q^2_s(x))}{2\,\kappa\,\lambda\,Y}\right] \ ,
\label{eq:dif:IIM}
\ee
which is the quadratic term in the exponent of Eq.(\ref{eq_T_fQ}), does
explictly violate the geometric scaling behaviour. We wish to emphasize here
that, as also shown in~\cite{Iancu:2003ge}, this violation seems required in
order to get an accurate description of the DIS data. Without it, even allowing
$\lambda_s$ to be an additional fitting paramter, one can not get a better
description of the DIS data. For further details on the importance of the
diffusion term see Ref.~\cite{Iancu:2003ge}. 

In this work, we wish to elaborate whether the violation of the geometric scaling
may come from gluon number fluctuations (Pomeron loops) and not from the
BK-equation. As we will see in the next sections, the fluctuations do indeed
give a similar violation of the geometric scaling and also lead to a better
description of the DIS data as compared to the case where the $T$-matrix shows
a geometric scaling behaviour.

\section{Beyond the mean field approximation} 
\label{sec:bmfa}
To go beyond the mean field approximation one has to include the effect of
discreteness and fluctuations of gluon
numbers~\cite{Mueller:2004se,Iancu:2004es}. After including fluctuations one
has to distinguish between the even-by-event amplitude and the averaged
(physical) amplitude. They can be explained by considering the evolution of a
proton from $y=0$ up to $y=Y$ which is probed by a dipole of size $r$, giving
the amplitude $\bar{T}(r,Y)$. The evolution of the proton is stochastic
and leads to random gluon number realizations inside the proton at $Y$,
corresponding to different events in an experiment. The physical amplitude,
$\bar{T}(r,Y)$, is then given by averaging over all possible gluon number
realizations/events, $\bar{T}(r,Y) = \langle T(r,Y)\rangle$, where $T(r,Y)$ is
the amplitude for the dipole $r$ scattering off a particular realization of
the evolved proton at $Y$. In the following we discuss the event-by-event
amplitude $T(r,Y)$ and the averaged amplitude $\bar{T}(r,Y)$.
\subsection{Event-by-event  scattering amplitude}
\label{subsec:se}
In a single scattering process, the mean field approximation breaks down when
the occupancy of gluons inside the evolved proton is low so that the
discreteness of the gluon number needs to be taken into account; the number of
gluons cannot be non-zero and less than one since it has to be discrete. When
including the discreteness effect, as compared to the results from the
BK-equation, the energy dependence of the saturation momentum changes
to~\cite{Mueller:2004se,Iancu:2004es}
\begin{equation}
Q_s^2(Y) = Q_0^2 \ \exp\!\left[\bar{\alpha_s}
\chi^{\prime}(\lambda_s) Y \left(1-\frac{\pi^2
    \chi''(\lambda_s)}{2(\Delta\rho)^2 \chi(\lambda_s)}\right)\right]
\label{eq:Qrbmf}
\end{equation}
and the piecewise, approximate, shape of the $T$-matrix at fixed coupling and
very high energy reads~\cite{Mueller:2004se,Iancu:2004es}

\begin{equation}
T(r,Y) = \left\{ \begin{array}{l@{\quad\mbox{for}\quad}l}
1 & \rho-\rho_s(Y) \ll 0
\vspace*{0.5cm}\\
N_1 \, [\rho-\rho_s(Y)]\,e^{\lambda_s [\rho-\rho_s(Y)]} & 0 < \rho-\rho_s(Y) <
\Delta\rho 
\vspace*{0.5cm}\\
N_2\, e^{-[\rho-\rho_s(Y)]} & \rho-\rho_s(Y) \gg \Delta\rho 
\end{array} \right.
\label{eq:ebe}
\end{equation}
where $N_1$ and $N_2$ are irrelevant constants and the front width is
$\Delta\rho \simeq (1/\lambda_s) \ln(1/\alpha^2)$.  The front width cannot be
larger than $\Delta\rho$ which is the distance when the amplitude decreases
from its maximal value $T \approx 1$ down to the value $T = {\cal
  O}(\alpha^2)$ where the discreteness of gluon numbers becomes important.
The width is formed via diffusion, $\rho-\rho_s(Y) \propto \sqrt{\alpha Y}$,
and it requires the rapidity $Y_F \simeq (\Delta\rho)^2/(2
\chi^{\prime\prime}(\lambda_s)\alpha)$ until it is completed. The
event-by-event amplitude given in Eq.(\ref{eq:ebe}), which is formed at $Y >
Y_F$, shows, approximately, geometric scaling: $T(r,Y) \approx
T(\rho-\rho_s(Y))$.

The main differences as compared to the mean-field results are: The exponent of the saturation scale in the
event-by-event amplitude, cf. Eq.(\ref{eq:Qrbmf}) and Eq.~(\ref{eq_Q_s}), is
decreased due to the discreteness of gluon numbers. Further the width of the
front of the event-by-event amplitude is fixed, $\Delta\rho$, instead of
increasing with rapidity as in Eq.(\ref{eq_T_fQ}).
%
%
\subsection{Physical scattering amplitude}
\label{ss:phys_amp}
Based on the relation between high-energy QCD evolution and reaction-diffusion
processes in statistical physics~\cite{Iancu:2004es}, the fluctuations in
gluon numbers are taken into account by averaging over all event-by-event
amplitudes,
\begin{equation}
\langle T((\rho-\rho_s(Y))\rangle = \int d\rho_s\ T(\rho-\rho_s(Y)) \
P(\rho_s(Y)-\langle\rho_s(Y)\rangle) \ \ ,
\label{av_gd}
\end{equation}
where the distribution of $\rho_s(Y)$ is, to a very good approximation, a
Gaussian~\cite{Marquet:2006xm}:
\begin{equation}
P(\rho _{s})\simeq 
\frac{1}{\sqrt{\pi \sigma ^{2}}}\exp \left[ -\frac{
\left( \rho _{s}-\langle \rho _{s}\rangle \right) ^{2}}{\sigma ^{2}}\right] \ .
\label{proba_gauss}
\end{equation}
The expectation value of the front position, $\langle \rho_s(Y)\rangle$,
increases with rapidity as $\langle \rho_s(Y)\rangle =\ln(Q^2_s(Y)/Q_0^2)$ at
high energy~\cite{Iancu:2004es}, with $Q_s(Y)$ given in Eq.~(\ref{eq:Qrbmf}).
The dispersion of the front at high energy increases linearly with rapidity,
\begin{equation}
\sigma^2 = 2 \left[\langle \rho_s^2 \rangle- \langle \rho_s \rangle^2\right] = D \, Y
\end{equation}
where $D$ is the diffusion coefficient, whose value is known only for
$\alpha \to 0$ (asymptotic energy)~\cite{Brunet:2005bz,Mueller:2004se}. 
%
Since the values of $D$ and the exponent $\lambda$ of the saturation scale,
$Q^2_s(x) =1\,\mbox{GeV}^2\, (x_0/x)^\lambda$, see Eq.~(\ref{eq:Qrbmf}), are
not known for finite energies, e.g. at HERA energy, in what follows we will
treat them as free parameters.

At very high energy, such that $\sigma^2 \gg 1$, the dispersion of the fronts
due to the gluon number fluctuations from event to event has large
consequences on $\langle T(r,Y)\rangle$: the geometric scaling of the single
events $T(\rho-\rho_s(Y))$, cf.  Eq.(\ref{eq:ebe}), is replaced by a new form
of scaling, known as diffusive scaling, namely, $\langle T(r,Y)\rangle$ is a
function of $(\rho-¸\langle\rho_s(Y)\rangle)/\sqrt{DY}$,
\be
\langle T(r,Y)\rangle = {\bar T}(r,Y) = {\bar T}\left(\frac{ \rho-\langle
    \rho_s(Y)\rangle }{\sqrt{DY}}\right) \ .
\label{eq:ds}
\ee
The diffusive scaling is expected to set in at $Y > Y_D =1/D$, which follows
from the requirement $\sigma^2 \gg 1$.

The goal of this work is to study whether the diffusive scaling behaviour of
the dipole-proton scattering amplitude in Eq.~(\ref{eq:ds}), which is caused
by gluon number fluctuations, may be present in the HERA data. 
As we will see in the next section, the fluctuations do improve the
description of the HERA data, indicating that the violation of geometric
scaling seems important for an accurate description of the data. We will
discuss whether the violation prefered by the DIS data is
due to the gluon number fluctuations, which lead to the diffusive scaling
$(\rho-\rho_s(Y))/\sqrt{DY}$, or due to the BK diffusion term, cf.
Eq.(\ref{eq:dif:IIM}), which corrects the geometric scaling in a similar way, 
namely, via $(\rho-\rho_s(Y))/\sqrt{2\,\bar{\alpha}\,\chi^{\prime\prime}\,Y}$.
\section{Numerical results}

Our fit includes the ZEUS data for the $F_2$ structure function, 
%
\be
 F_2(x,Q^2) = \frac{Q^2}{4\pi^2\alpha_{em}}\ (\sigma_T(x,Q^2) +
 \sigma_L(x,Q^2)) \ , \quad \quad \sigma_{T,L}(x,Q^2) = \int \ dz\,d^2r \ |\psi_{T,L}(z,r,Q^2)|^2 \ \sigma_{dip}(x,r)
\label{eq:Fst}
\ee
in the kinematical range $x \leq 10^{-2}$ and $ 0.045 \,\mbox{GeV}^2 < Q^2 <
50\,\mbox{GeV}^2$ (see also~\cite{Iancu:2003ge} for more discussions on the range). 
The upper limit on $Q^2$ has been chosen large enough to include a large
amount of ``perturbative'' data points, but low enough in order to justify the
use of the BFKL dynamics, rather than DGLAP evolution. We use in our fit the
same photon wave functions $\psi_{T,L}$ as in Ref.\cite{GolecBiernat:1998js},
which are computable in QED, and three quarks with equal mass, $m_q =
140\,\mbox{MeV}$. We have considered only the ZEUS data because there is a
mismatch between the H1 and ZEUS with regard to the data normalization and
since only ZEUS has data also in the low $Q^2$ region, i.e., in the saturation
region. To fix the parameters we minimize $\chi^2 = \sum_i
(\mbox{model}(i,p_1,...,p_n) -F_2(i))^2/(\mbox{error}(i))^2$, where the sum
goes over the data points, $p_1, ..., p_n$ denote the parameters to be
found, $F_2(i)$ the experimental results for the $F_2$ structure function, and
for the error of $F_2$, i.e., $(\mbox{error}(i))^2$, we use the systematic error squared plus the
statistical error squared.

The interesting ingredient for us in Eq.~(\ref{eq:Fst}) is the dipole-proton
cross section, $\sigma_{dip} = 2 \pi R^2 \ \langle T(r,x) \rangle$, with $2
\pi R^2$ being the outcome of the integration over the impact parameter.  We
will use different ans\"atze for the event-by-event amplitude, $T(r,x)$, and
the physical amplitude, $\langle T(r,x) \rangle$, is obtained according to the
rules outlined in section~\ref{ss:phys_amp}. (We wish to note that the
ans\"atze for $T(r,x)$, which are derived/motivated based on perturbative QCD,
are used to describe also the low virtuality data, $Q^2 \leq 1\,\mbox{GeV}^2$,
in the fit to the HERA data. In this region non-perturbative
physics~\cite{Ewerz:2006vd} is involved which is only approximately given by
our ans\"atze.)  In $\sigma_{dip}$ we will use the event-by-event amplitude
and the physical amplitude in order to study the effects of gluon number
fluctuations. In the case of $T(r,x)$ there are three free parameters which
will be fixed by fitting the HERA data: $R$ (``radius of the proton'') and
$x_0$ and $\lambda$ coming via the saturation momentum $Q^2_s(x)
=1\,\mbox{GeV}^2\, (x_0/x)^\lambda$. In the case of the averaged (physical)
amplitude, $\langle T(r,x) \rangle$, there is another free parameter, the
diffusion coefficient $D$.

%
\begin{itemize}
\item Golec-Biernat, W\"usthoff (GBW) model~\cite{GolecBiernat:1998js}: \\

The GBW model
\be
T^{GBW}(r,x) = 1- \exp\left[-\frac{r^2\,Q^2_s(x)}{4}\right] \ ,
\label{eq:GBW}
\ee
is one of the most simple models which shows geometric scaling, $T(r,x) =
T(r^2\,Q^2_s(x))$, and leads to a quite successful description of the HERA
data, as can be seen from Figs.~\ref{Fig_GBW1},~\ref{Fig_GBW2} and the
$\chi^2$ (error) in Table~\ref{tab_GBW} (denoted by GBW). It is nice to see
that the value of the saturation exponent, $\lambda \simeq 0.285$, which is
found by fitting the HERA data with the GBW model, comes out close to the
theoretical NLO results for $\lambda$~\cite{Triantafyllopoulos:2002nz}.

Now, using the GBW model as an event-by-event amplitude, we include the effect
of gluon number fluctuations by averaging over all events via
Eq.~(\ref{av_gd}).  The resulting $\langle T^{GBW}(r,x) \rangle$, which breaks
the geometric scaling, leads to a relatively much better description of the
HERA data, as can be seen from the comparision of the $\chi^2$ values and the
two lines in Figs.~\ref{Fig_GBW1},~\ref{Fig_GBW2}. The large improvement after
including fluctuations seems to indicate that violations of geometric scaling,
and probably even gluon number fluctuations, are implicated in the HERA data.

It is important to note that the values of the fitting parameters come out
reasonable also after including the gluon number fluctuations. The value of
$\lambda$ becomes smaller after including fluctuations which is in agreement
with theoretical expectations, as can be seen from the comparison of
Eq.~(\ref{eq_Q_s}) with Eq.~(\ref{eq:Qrbmf}). Furthermore, the value of the
diffusion coefficient $D$ is sizeable, and is surprisingly close to the values
which have been found numerically by solving the ($1+1$) dimensional toy
model~\cite{Iancu:2006jw} and the approximate QCD evolution
equations~\cite{Soyez:2005ha} (they represent an approximation of the Pomeron
loop equations~\cite{Mueller:2005ut,Iancu:2004iy,Kovner:2005nq}) in the fixed
coupling case. Note also that the radius of the proton, $R$, increases
somewhat and $x_0$ becomes smaller, meaning that $Q_s <1\,\mbox{GeV}$ up to
$x\simeq 10^{-6}$, due to fluctuations. Also the reasonable values of the
parameters, especially the sizeable value of $D$ yielding $Y_D = 1/D \simeq
2.5$, in addition to the better description of the HERA data after including
fluctuations, seem to be in favour of an implication of gluon number
fluctuations in the HERA data.
\begin{table}
\begin{center}
\begin{tabular}{r@{\quad}||c@{\quad}|c@{\quad}|c@{\quad}|c@{\quad}|c@{\quad}|c@{\quad}|}
model/parameters & \quad $\chi^2$ & \quad $\chi^2/\mbox{d.o.f}$ & \quad $x_0$
($\times 10^{-4}$) & \quad $\lambda$ &
\quad $R$(fm) & \quad $D$ \\ [0.5ex] \hline \hline
$T^{\mbox{\footnotesize GBW}}$ & \quad 266.22 & \quad 1.74 & \quad 4.11 & \quad 0.285 & \quad 0.594 &
\quad 0 \\ \hline
$\langle T^{\mbox{\footnotesize GBW}} \rangle$ & \quad 173.39 & \quad 1.14 & \quad 0.0546 & \quad
0.225 & \quad 0.712 & \quad 0.397 \\ \hline
\end{tabular}
\caption{GBW model: The parameters of the
  event-by-event ($2$ line) and of the physical ($3$ line) amplitude.}
\label{tab_GBW}
\end{center}
\end{table}
\begin{figure}[h!]
\setlength{\unitlength}{1.cm}
\begin{center}
\epsfig{file=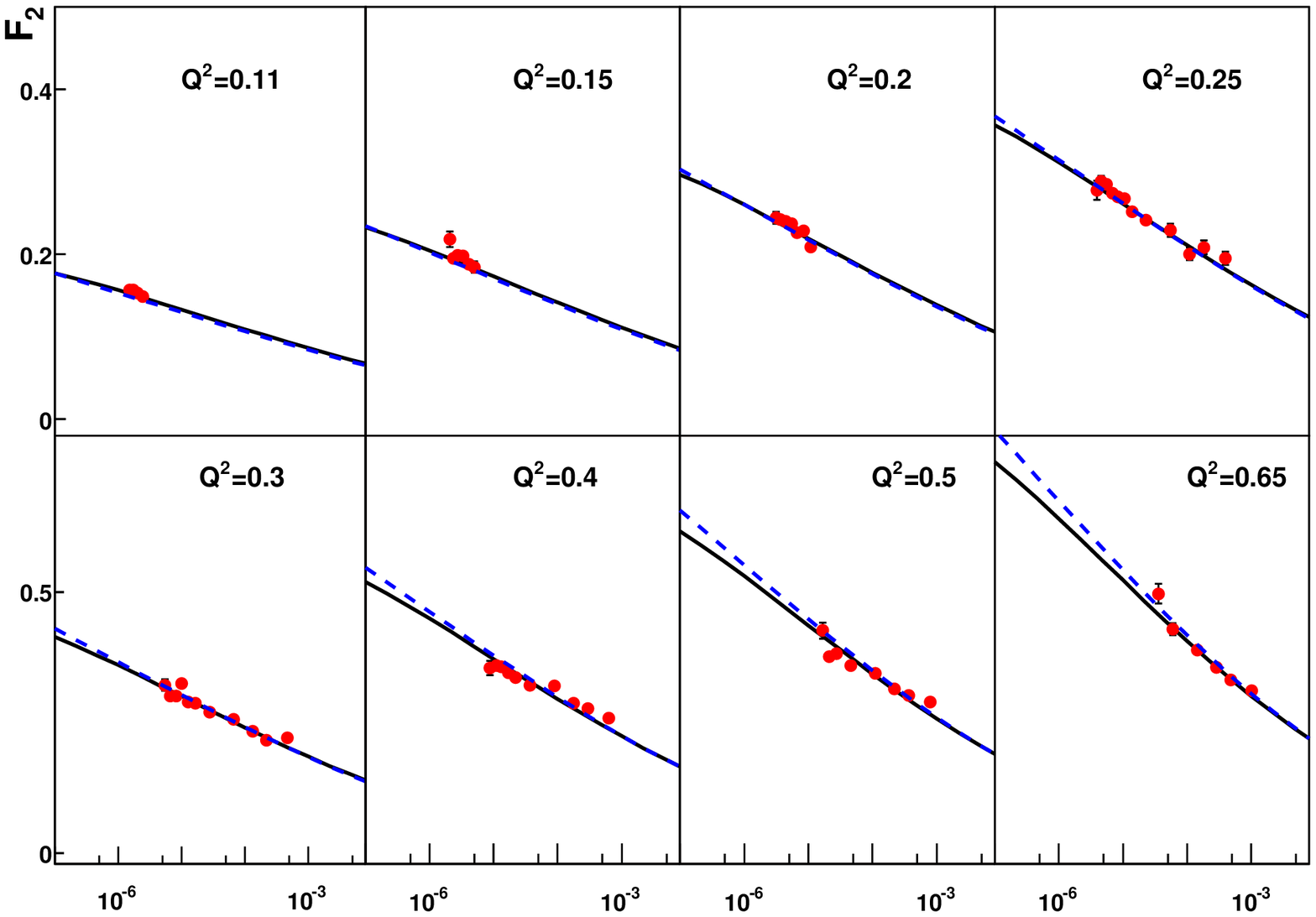, width=14.5cm} \\
\vspace*{0.3cm}
\epsfig{file=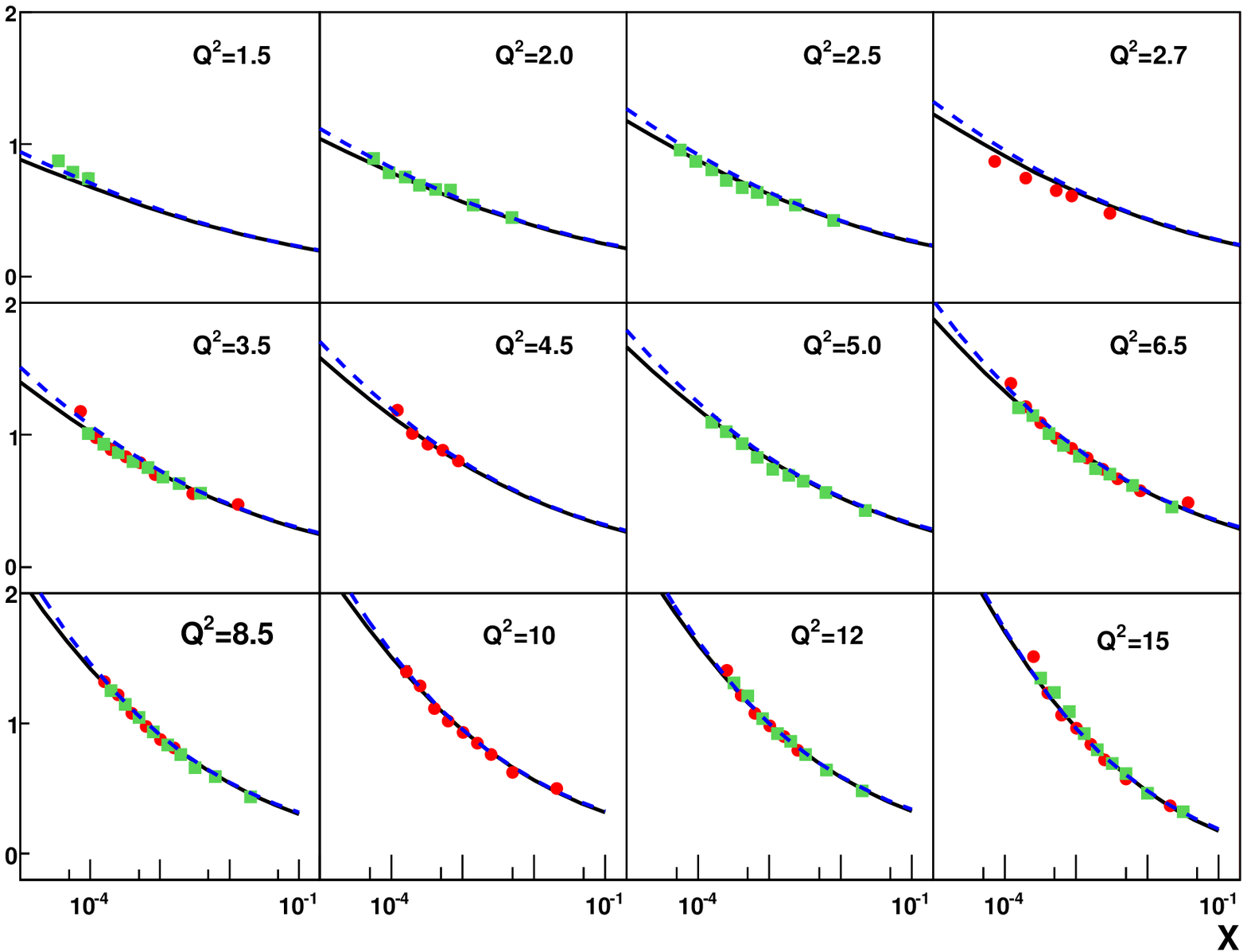, width=14.9cm} 
\end{center}
\caption{The $F_2$ structure function versus $x$ at different values of $Q^2$.
  The experimental points are the latest published data from the H1 (squares)
  and ZEUS (dots) collaborations~\cite{Breitweg:2000yn+X}. The solid line represents the
  result of the averaged GBW fit and the dashed line the result of the GBW fit
  to the ZEUS data for $x \leq 10^{-2}$ and $ 0.045 \,\mbox{GeV}^2 < Q^2 <
  50\,\mbox{GeV}^2$. The data points at lowest $Q^2$ values, $0.045$, $0.065$
  and $0.085\,\mbox{GeV}^2$, are not shown here although they are included in
  the fits.}
\label{Fig_GBW1}
\end{figure}
\begin{figure}[h!]
\setlength{\unitlength}{1.cm}
\begin{center}
\epsfig{file=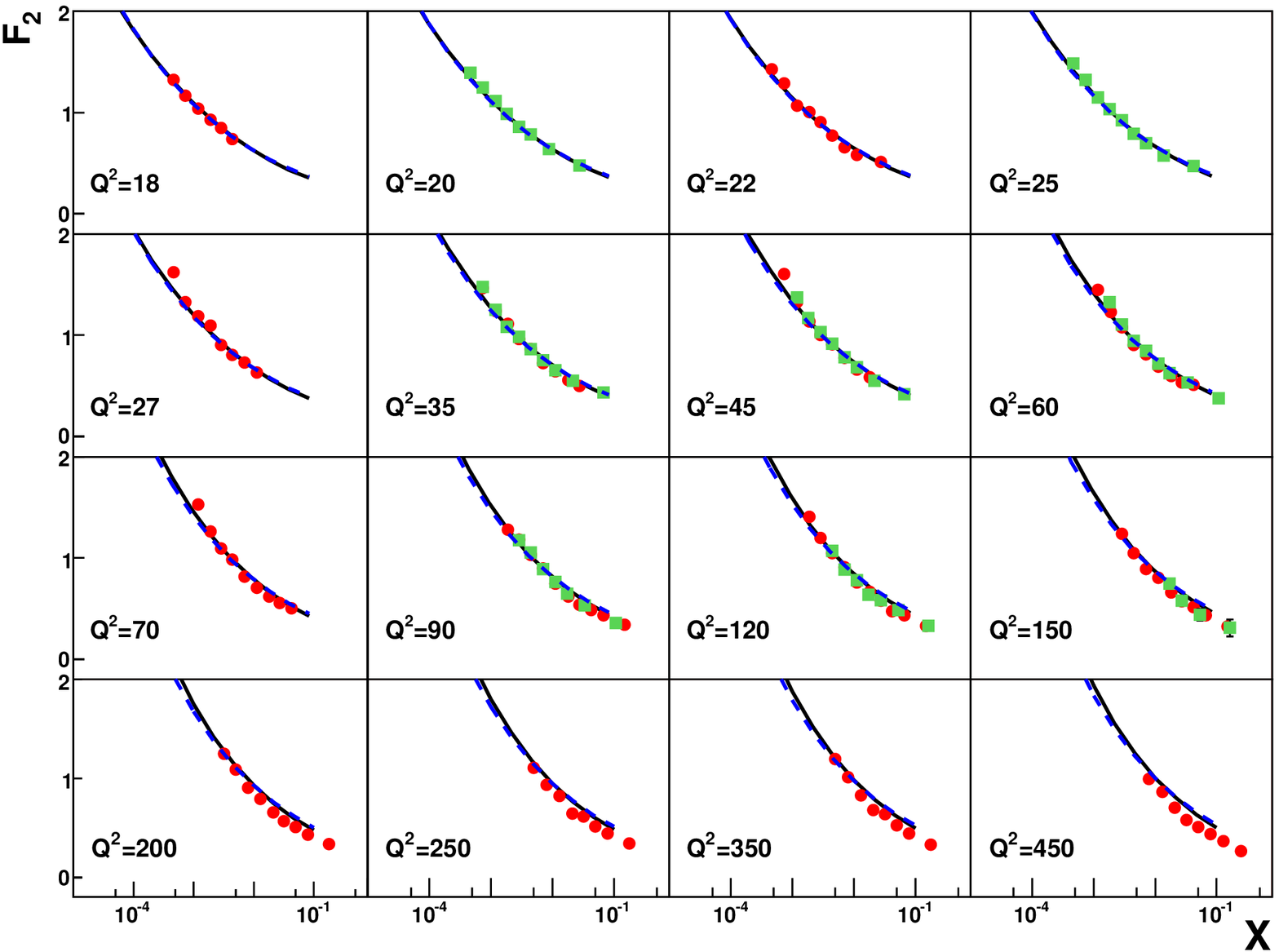, width=16cm}
\end{center}
\caption{The same as in Fig.~\ref{Fig_GBW1}, but for larger values of $Q^2$.
  Note that we show in this figure our results up the highest $Q^2$ although
  our fit is performed inluding only the data for $Q^2 < 50\,\mbox{GeV}^2$.}
\label{Fig_GBW2}
\end{figure}
\item Iancu, Itakura, Munier (IIM) model~\cite{Iancu:2003ge}, other models and
  a model-independent approach: \\

  The IIM model given in Eq.~(\ref{eq:IIM}) includes the BK-diffusion term,
  $\ln(4/r^2\,Q_s^2)/\sqrt{2\,\kappa\,\lambda\,Y}$, which explicitly violates
  the geometric scaling. It has been shown in~\cite{Iancu:2003ge} that this
  violation does noticeably improve the description of the HERA data in
  comparison with the GBW model, as can be seen from the much smaller $\chi^2$
  value in the IIM case in Table~\ref{tab_IIM} (we always use $N_0=0.5$ in the
  IIM model). In Ref.~\cite{Iancu:2003ge} has been further shown that without
  the BK-diffusion term, although allowing for an additional free parameter
  $\lambda_s$ (one parameter more than in the GBW model), the
  $\chi^2/\mbox{d.o.f}$ value does not improve and is close to the GBW value.

\begin{table}[htp]
\begin{center}
\begin{tabular}{r@{\quad}||c@{\quad}|c@{\quad}|c@{\quad}|c@{\quad}|c@{\quad}|c@{\quad}|}
model/parameters & \quad $\chi^2$ & \quad $\chi^2/\mbox{d.o.f}$ & \quad $x_0$
($\times 10^{-4}$) & \quad $\lambda$ &
\quad $R$(fm) & \quad $D$ \\ [0.5ex] \hline \hline
$T^{\mbox{\footnotesize IIM}}$ & \quad 150.45 & \quad 0.983 & \quad 0.5379 & \quad 0.252 & \quad 0.709 &
\quad 0 \\ \hline
$\langle T^{\mbox{\footnotesize IIM}} \rangle$ & \quad 122.62 & \quad 0.807 & \quad 0.0095 & \quad
0.198 & \quad 0.812 & \quad 0.325 \\ \hline
\end{tabular}
\caption{IIM model: The parameters of the
  event-by-event ($2$ line) and of the physical ($3$ line) amplitude.}
\label{tab_IIM}
\end{center}
\end{table}

Note that the GBW model only after including gluon number fluctuations gives a
$\chi^2/\mbox{d.o.f}$ value which is comparable with the IIM one. This may
mean that the violation of the geometric scaling is favoured by the HERA data.
The violation may come from the gluon number fluctuations or from the
BK-diffusion term.

In the case of the IIM model, after including fluctuations, we can give an
analytic expression for the physical amplitude
\bea 
\langle T^{\mbox{\footnotesize IIM}}(r,Y) \rangle &=& \frac{N_0}{2 \sigma} \left[ 
\sigma\, \mbox{Erfc}\left(\frac{\ln\frac{4}{r^2Q_s^2}}{\sigma} \right)
-
\frac{\mbox{Exp}\left(\frac{-\frac{a}{4\sigma^2}\,\ln^2(b^2\,r^2Q_s^2)}{\frac{1}{\sigma^2}+\frac{a}{4}}\right)}{\sqrt{\frac{1}{\sigma^2}+\frac{a}{4}}}
\times 
\mbox{Erfc}\left(\frac{\frac{a \ln(4b^2)}{4} + \frac{1}{\sigma^2}\,\ln(\frac{4}{r^2Q_s^2})}{\sqrt{\frac{1}{\sigma^2}+\frac{a}{4}}}
\right) \right.
\nonumber \\
&&\hspace*{-0.8cm}+ \left. \frac{1}{\sqrt{\frac{1}{2\kappa \lambda Y}+\frac{1}{\sigma^2}}} \ 
\left(1 + \mbox{Erf}\left(\frac{-\frac{\lambda_s}{2} +
      \frac{1}{\sigma^2}\,\ln(\frac{4}{r^2Q_s^2})}{\sqrt{\frac{1}{2\kappa
          \lambda Y}+\frac{1}{\sigma^2}}}\right)\right) 
\times \ 
\mbox{Exp}\left(- \frac{\left[
    \frac{\ln^2(\frac{4}{r^2Q_s^2})}{2\kappa \lambda Y \sigma^2} -
    \frac{\lambda_s^2}{4} + \lambda_s \frac{\ln(\frac{4}{r^2Q_s^2})}{\sigma^4}
  \right]}{\sqrt{\frac{1}{2\kappa \lambda Y}+\frac{1}{\sigma^2}}} 
\right) \right] \ , \nonumber
\eea
which can be used in phenomenological applications, where $\mbox{Erfc(x)}$ is
the complementary error function. Also in the IIM case fluctuations do improve
the description of the HERA data, however not much, as can be seen from the
comparable $\chi^2/\mbox{d.o.f}$ values for $T^{\mbox{\footnotesize IIM}}$ and
$\langle T^{\mbox{\footnotesize IIM}}\rangle$ in Table~\ref{tab_IIM}. This is
so because the IIM model does already contain the geometric scaling violations
via the BK-diffusion term, $\ln(4/r^2Q_s^2)/\sqrt{2\kappa \lambda Y}$, and
describes accuratelly the HERA data, before including fluctuations. However,
note that the diffusion coefficients in case of fluctuations and the
BK-diffusion term are quite different, namely, $D=0.325$ and
$2\,\kappa\,\lambda \simeq 3.9$, respectivelly.

After including fluctuations, the parameters in the GBW and the IIM case are
close to each other. Apart from the fact that similar values for $D$ are found
in numerical simulations of evolution
equations~\cite{Soyez:2005ha,Iancu:2006jw} and the decrease of $\lambda$ due
to fluctuations is theoretically expected, at least at high energy, the
parameters $\lambda$ and $D$ also seem to be quite model-independent. Indeed,
similar values for $\lambda$ and $D$ would come out also if one uses a
model~\footnote{Such a model would be for instance the IIM model with the
  diffusion variable $\ln(4/r^2\,Q_s^2)/\sqrt{2\,\kappa\,\lambda\,Y}$ replaced
  by $\ln(4/r^2\,Q_s^2) (1-\lambda_s)/\sqrt{\Delta\rho}$, such that the new
  model interpolates between the three regions of Eq.(\ref{eq:ebe}) and shows
  the geometric scaling behaviour. The constant $\Delta\rho$ is given by
  Eq.(\ref{eq:ebe}). We use in $\Delta\rho$ a small value for $\alpha$,
  $\alpha=1/15$, which is the value required such that the exponent of $Q_s^2$
  in Eq.(\ref{eq_Q_s}) agrees with experimental or NLO results, $\lambda
  \simeq 0.3$. With this input, we find for $R=0.8\,\mbox{fm}$, the following
  results: $\lambda=0.235$ and $D=0.58$.} as suggested by the thoeretical
findings at high energy as given in Eq.~(\ref{eq:ebe}), for reasonable values
of the proton radius, $R \simeq 0.7 - 0.8$\,fm. Moreover, the similar value of
$\lambda$ coming out of the different models is also supported by the
following model-independent approach: In case fluctuation are important in the
range of HERA data, one finds the diffusive scaling
behaviour~\cite{Hatta:2006hs}, i.e., $\sigma^{\gamma^*p}/\sqrt{DY}$ is a
function of $\ln(1/r^2\,Q_s^2)/\sqrt{DY}$.  Defining a ``quality factor''
${\cal O}(\lambda, x_0, D)$ as done in~\cite{Gelis:2006bs}, which test the
quality of this diffusive scaling in HERA data, we find by minimazing it
$\lambda = 0.215$, at least for the input-values $0.01\leq D \leq 0.7$ which
we have investigated.

The seemingly model-independent values of the parameters $\lambda$ and $D$,
their agreement with numerically found values, and the improvement of the
description of the HERA data in all models after including fluctuations, seem
to tell us that gluon number fluctuations are relevent in the range of HERA
data. However, since in the case of the IIM model the fluctuations do not
improve much the description of the HERA data, one may conclude that the
BK-equation alone should describe the HERA data and that fluctuations are
neglegible in the energy range of the HERA data. The intention of this work is
to illustrate the possibility that fluctuations may be present in the HERA
data.  

The situation would become more clear at even higher collision energies as
compared to the HERA energy. With growing $Y$, according to the BK-equation
the window for the geometric scaling behaviour would increase, and the scaling
violating term would become less important. On the other side, the small-$x$
dynamics including gluon number fluctuations leads to a more clear diffusive
scaling behaviour with increasing $Y$. The forthcoming LHC may tell us more
whether geometric or diffusive scaling is more appropriate for the description
of the observables in the LHC energy range.  

Throughout this work the coupling is kept fixed. As mentioned above, the
($1+1$) dimensional model in~\cite{Iancu:2006jw}, which accomodates Pomeron
loops, gives similar values for $D$ as our analysis for a fixed coupling.
However, it has been recently shown that if allowing the coupling to run in
this toy model~\cite{Dumitru:2007ew} then the effects of gluon number
fluctuations can be neglected up to energies far beyond the HERA and LHC
energies. We plan to extend our work by the running coupling in order to see
whether the HERA data can tell something about the running coupling and
whether the prediction of the toy model remains valid also in the QCD case.

\begin{acknowledgments}
  We are very thankful to Stephane Munier for providing us with the codes
  which he has used for the analysis of the HERA data with the IIM model.
  We are grateful to Edmond Iancu for valuable discussions.  The authors
  acknowledge financial support by the Deutsche Forschungsgemeinschaft under
  contract Sh 92/2-1.
\end{acknowledgments}
%

\end{itemize}

\end{document}